\documentclass[pdflatex,sn-basic]{sn-jnl}

\usepackage{graphicx}%
\usepackage{multirow}%
\usepackage{amsmath,amssymb,amsfonts}%
\usepackage{amsthm}%
\usepackage{mathrsfs}%
\usepackage[title]{appendix}%
\usepackage[dvipsnames,usenames]{xcolor}%
\usepackage{textcomp}%
\usepackage{manyfoot}%
\usepackage{booktabs}%
\usepackage{algorithm}%
\usepackage{algorithmicx}%
\usepackage{algpseudocode}%
\usepackage{listings}%
\usepackage{subcaption}%
\usepackage{dsfont}
\usepackage{booktabs,tabularx}

\theoremstyle{thmstyleone}%
\theoremstyle{thmstyletwo}%

\theoremstyle{thmstylethree}%
\newcommand{\ba}{\boldsymbol{a}}

\newcommand{\br}{\boldsymbol{r}}
\newcommand{\bp}{\boldsymbol{p}}
\newcommand{\bw}{\boldsymbol{w}}

\newcommand{\bv}{\boldsymbol{v}}

\renewcommand{\bf}{\boldsymbol{f}}

\newcommand{\bT}{\boldsymbol{T}}
\newcommand{\bA}{\boldsymbol{A}}

\newcommand{\bV}{\boldsymbol{V}}

\newcommand{\bR}{\boldsymbol{R}}
\newcommand{\bC}{\boldsymbol{C}}
\newcommand{\bD}{\boldsymbol{D}}
\newcommand{\bS}{\boldsymbol{S}}
\newcommand{\bZ}{\boldsymbol{Z}}
\newcommand{\bLambda}{\boldsymbol{\Lambda}}
\newcommand{\bSigma}{\boldsymbol{\Sigma}}

\newcommand{\real}[1]{\mathds{R}^{#1}}

\newcommand{\calW}{\mathcal{W}}

\newcommand{\calP}{\mathcal{P}}
\newcommand{\calA}{\mathcal{A}}

\DeclareMathOperator{\diag}{diag}

\newcommand{\T}{\textnormal{\textsf{T}}}

\newcommand{\bone}{\boldsymbol{1}}
\newcommand{\bzero}{\boldsymbol{0}}

\newcommand{\brc}[1]{\left( #1 \right)}
\newcommand{\sqbrc}[1]{\left[ #1 \right]}
\newcommand{\figbrc}[1]{\left\{ #1 \right\} }

\renewcommand{\d}{\; \mathrm{d}}

\renewcommand{\arg}{\mathrm{arg}}

\begin{document}

\title[Correlation Structures and Regime Shifts in Nordic Stock Markets]{Correlation Structures and Regime Shifts in Nordic Stock Markets}


\author*[]{\fnm{Maksym A.} \sur{Girnyk}}\email{82735@student.hhs.se}



\affil[]{\orgdiv{Executive MBA}, \orgname{Stockholm School of Economics}, \city{Stockholm}, \country{Sweden}}




\abstract{Financial markets are complex adaptive systems characterized by collective behavior and abrupt regime shifts, particularly during crises. This paper studies time-varying dependencies in Nordic equity markets and examines whether correlation-eigenstructure dynamics can be exploited for regime-aware portfolio construction. Using two decades of daily data for the OMXS30, OMXC20, and OMXH25 universes, pronounced regime dependence in rolling correlation matrices is documented: crisis episodes are characterized by sharp increases in the leading eigenvalue and counter-cyclical behavior in the second eigenvalue. Eigenportfolio regressions further support a market-factor interpretation of the dominant eigenmode. Building on these findings, an adaptive portfolio allocation framework is proposed, combining correlation-matrix cleaning, an eigenvalue-ratio crisis indicator and long-only minimum-variance optimization with constraints that bound exposures to dominant eigenmodes. Backtesting results indicate improved downside protection and risk-adjusted performance during stress regimes, while remaining competitive with state-of-the-art benchmarks in tranquil periods.}





\maketitle

\section{Introduction}
\label{sec:intro}

Financial markets are characterized by complex interactions among assets,
institutions, and macroeconomic forces. These interactions evolve over time and
often give rise to episodes of market stress marked by sharp increases in
co-movement, volatility clustering, and breakdowns in diversification. Such
features pose significant challenges for traditional equilibrium-based models
of asset pricing and portfolio choice, including the Efficient Markets
Hypothesis, the Capital Asset Pricing Model and representative-agent
frameworks~\citep{campbellLo1997econometrics,osullivan2018capital,lucas1978asset}.
Understanding how dependence structures change across regimes is essential for
characterizing systemic risk, identifying periods of market stress and
designing portfolios that remain robust when diversification fails. 

The present paper investigates how regime-indicative correlation dynamics
in \emph{Nordic equity markets} can be used to improve portfolio performance, especially during
periods of market stress. Building on the recent findings of~\citet{molero2025random}, who document 
the defensive properties of the second eigenvalue of the return correlation matrix and its associated 
eigenvector in high-volatility regimes, I consider a regime-aware allocation perspective. The central 
hypothesis is that shifts in the spectral structure of the correlation matrix convey timely information 
about emerging crises and can therefore be exploited to construct more defensive portfolio allocations.

The structure and time variation of financial correlations are examined, focusing on the constituent 
equities of Nordic stock markets, comprising the OMXS30 (Sweden), OMXC20 (Denmark) and OMXH25 (Finland) indices. 
Nordic markets offer an informative empirical setting: they are relatively small open economies,
highly integrated through trade and finance and strongly exposed to global
macroeconomic factors. Over the past two decades, these markets have been exposed to a series of salient 
stress episodes~\citep{diamond2024politics}---including the 2008 Global Financial Crisis, the 2011--12 Eurozone turmoil, 
the 2015--16 China's slowdown, commodity downturn and refugee crisis, the 2020 Covid-19 pandemic shock and the 2022 illegal russian 
invasion of Ukraine with its ensuing inflationary pressures and energy-price surge---which resulted in pronounced 
periods of variation in correlation regimes and cross-asset co-movement.

The analysis proceeds in several steps. First, I examine the eigenvalues 
of the correlation matrix to identify structural changes in
market-wide, sectoral and defensive components. Particular attention is given
to the counter-cyclical behavior of subleading eigenvalues recently documented
by~\citet{molero2025random}. Second, I conduct an eigenportfolio beta analysis based on univariate ordinary 
least squares regressions and estimate corresponding eigenportfolio 
betas. Third, I exploit the above insights to construct a
crisis-aware portfolio strategy. Namely, I propose an adaptive portfolio
approach that adjusts allocation to prevailing market regimes and benchmark 
this strategy against capable state-of-the-art approaches by evaluating their risk-adjusted performance. 

At a high level, my findings indicate that spectral features of Nordic equity correlation 
matrices contain actionable information about market regimes which can be translated into 
improved portfolio decisions. I document pronounced regime variation in the leading and 
subleading eigenvalues of Nordic correlation matrices, consistent with elevated market-wide 
co-movement during stress episodes and comparatively defensive behavior in lower-order components. 
I furthermore show that the dominant eigenmode closely tracks the market return, 
supporting a market-factor interpretation of the leading eigenvector. 
Finally, I show that the proposed regime-aware allocation scheme, which exploits 
spectral signals to adapt exposures across tranquil and stressed periods, tends to 
outperform established uninformed benchmarks in backtests on Nordic markets.

This paper contributes to the literature in four ways: 
(i) empirically, I document crisis-time dynamics of eigenvalues and eigenvectors in Nordic equity markets, 
providing new evidence on regime-dependent correlation structure and counter-cyclical behavior in leading eigenmodes;
(ii) methodologically, I introduce a correlation-based regime indicator suitable for determining the market state;
(iii) I further show that the principal eigenportfolio admits a Sharpe-style market-factor interpretation;
(iv) given the above findings, I propose a regime-aware defensive portfolio strategy that adapts to the 
prevailing market state using the regime indicator and compare it with established 
methods, thereby positioning the results within the broader literature on data-driven portfolio design.

The remainder of the paper is structured as follows. 
Section~\ref{sec:review} reviews the relevant literature on correlation dynamics, eigenvalue methods and Nordic financial markets under consideration. 
Section~\ref{sec:data} describes the dataset and outlines the institutional context. 
Section~\ref{sec:method} describes the modeling and analytical framework, outlines the tools employed and the performance evaluation protocol, as well as specifies the baseline portfolio strategies used for comparison.
Section~\ref{sec:portfolio_construction} introduces the proposed market-regime indicator and describes the associated regime-aware portfolio construction procedure.
Section~\ref{sec:results} presents the empirical results, compares portfolio performance across strategies and provides a discussion of the findings.
Section~\ref{sec:conclusions} concludes the paper and outlines directions for future research.

The following notation is used throughout the paper. Upper-case and lower-case boldface letters (e.g., $\bA$ and $\ba$) denote matrices and column vectors, respectively, while calligraphic letters (e.g., $\calA$) denote sets. Moreover, $\bzero$ and $\bone$ denote the vector of zeros and ones, respectively. The operator $\brc{\cdot}^{\T}$ denotes transpose. The mapping $\diag\brc{\bA}$ returns the diagonal matrix formed from the diagonal entries of $\bA$, while $\diag\brc{\ba}$ denotes the diagonal matrix with the elements of $\ba$ on its diagonal. Moreover, $\sqbrc{\bA}{i,j}$ denotes the $(i,j)$-th entry of matrix $\bA$. The functionals $\mathrm{var}\figbrc{\cdot}$, and $\mathrm{cov}\figbrc{\cdot,\cdot}$ denote the variance and covariance of their arguments, respectively. Finally, $z\figbrc{\cdot}$ denotes the $z$-score operator, viz., centering its argument by subtracting the mean and scaling by the corresponding standard deviation.

\section{Literature Review}
\label{sec:review}

Financial markets exhibit complex and time-varying interdependencies, motivating 
the widespread use of correlation matrices to summarize cross-asset return relationships. 
These matrices are typically constructed using the Pearson correlation coefficient, 
which quantifies linear association between asset returns and is among the most 
widely applied dependence measures. 
A large body of empirical work documents that correlations rise sharply during periods 
of market stress, reducing diversification opportunities and amplifying
systemic risk~\citep{longin2001extreme,forbes2002no,ang2002international,billio2012econometric}.
Since the seminal contribution of~\citet{markowitz1952portfolio}, modern portfolio theory 
has emphasized the role of cross-asset dependence in diversification and optimal allocation. 
These patterns are closely linked to market integration, contagion and the
transmission of global shocks across countries and sectors
\citep{bekaert2003market,diebold2012better}. 

Principal component analysis (PCA) provides a natural way to summarize high-dimensional
dependence structures~\citep[see, e.g.,][]{jolliffe2011principal}. For instance, the first principal component 
captures market-wide co-movement,
while subsequent components reflect sectoral, regional, or style-based
dynamics~\citep{king1966market, meyers1973re}. These tools have become standard for studying 
factor structure, spillovers and time-varying market integration~\citep{diebold2012better,molero2025random}.

A complementary body of literature applies Random Matrix Theory (RMT)~\citep[see, e.g.,][]{bouchaud2003theory} 
to separate signal from noise in large-dimensional correlation matrices. The pioneering
studies of~\citet{laloux1999noise} and~\citet{plerou2002random} demonstrate that 
the bulk of eigenvalues of empirical return correlation matrices lies within the 
Marchenko--Pastur (MP) bounds~\citep{marchenko1967distribution}, indicating that 
most modes are largely driven by finite-sample noise. In contrast, only a small 
number of eigenvalues appear as persistent outliers relative to the MP benchmark 
and can be interpreted as systematic components. These studies further show that 
the dominant eigenvalue is typically associated with a market-wide factor, whereas 
intermediate outliers may encode sectoral, regional, or other structured sources of co-movement. 
Subsequent research emphasizes the role of RMT in tracking structural changes in dependence, 
especially during crises when the leading eigenvalue increases sharply and subleading 
eigenmodes may exhibit defensive behavior~\citep{james2022financial,molero2025random,dominguez2025correlation}. 
These findings align with a broader literature showing regime dependence in correlation structure,
contagion and volatility clustering~\citep{ang2002international, forbes2002no, cont2001empirical}.

Although RMT-based techniques have been connected to portfolio optimization primarily 
via eigenvalue cleaning~\citep{laloux1999noise, sharifi2004random}, comparatively 
little work has focused on exploiting the temporal dynamics of the significant 
eigenvalues per se as signals for portfolio allocation. Existing PCA and factor-based strategies
\citep{connor1986performance,fan2013large} provide dimensionality reduction but
do not explicitly exploit the regime dependence of the eigenstructure. Meanwhile, 
recent studies suggest that subleading eigenmodes may move inversely to the market 
in crises ~\citep{nobi2013random,molero2025random}. This counter-cyclical behavior motivates 
their use for defensive portfolio allocation under elevated volatility.

The present study focuses on Nordic equity markets, which provide a salient empirical
setting due to their strong exposure to global shocks and repeated episodes of
market stress~\citep{imf2013nordic, farelius2025financial}. I examine how the 
eigenvalue spectrum evolves across crisis episodes and assess whether these dynamics 
can be leveraged to inform portfolio construction.

Existing relevant literature includes PCA-based measures of market tightness, such as 
the absorption ratio of \citet{kritzman2010principal}, and related approaches that 
interpret time variation in the variance explained by the leading principal components 
as changes in market integration \citep{volosovych2013learning}. Complementary streams 
develop dependence-based early-warning indicators using graph-theoretic and information-theoretic 
summaries, including eigenvector-centrality measures of correlation networks \citep{ciciretti2025early} 
and entropy diagnostics designed to capture turbulence-induced changes in market dynamics \citep{chakraborti2020phase, olbrys2022regularity}. RMT-based contributions further motivate spectral 
monitoring by linking stress to systematic departures of the empirical eigenvalue distribution from 
MP benchmarks, thereby separating structured components from sampling noise.

Within this broader agenda, several studies document that the correlation eigenstructure itself 
evolves markedly across crises. Rolling-window analyses report crisis-time strengthening of collective 
behavior reflected in the leading eigenvalue and its associated eigenvector \citep{junior2012correlation, song2011evolution}, 
while earlier evidence suggests that subdominant eigenvalues can carry additional information about market 
reactions and recovery following major events \citep{sharkasi2006reaction}. Recent work extends this perspective 
by emphasizing that subleading eigenmodes may exhibit counter-cyclical, potentially defensive behavior in 
high-volatility regimes \citep{molero2025random}. In parallel, regime-aware portfolio design has primarily 
relied on latent-state models or alternative dependence proxies, including dynamic optimization based on hidden Markov models \citep{nystrup2018dynamic} and allocations informed by network centrality \citep{ciciretti2023building}. 
Against this backdrop, the present study focuses on Nordic equity markets and investigates whether 
the time variation of significant correlation eigenvalues, particularly subleading eigenmodes, can be 
operationalized as a crisis indicator and translated into an implementable regime-aware allocation 
rule that adapts systematically across tranquil and stress periods.

\section{Data and Background}
\label{sec:data}

\subsection{Dataset}
This study uses daily closing prices for the constituent stocks of three major
Nordic equity indices: the OMX Stockholm 30 (OMXS30, Sweden), the OMX Copenhagen
20 (OMXC20, Denmark), and the OMX Helsinki 25 (OMXH25, Finland).  The dataset
covers the period from January 2006 to September 2025 and is retrieved from
\citet{GoogleFinance2025}, where prices are adjusted for splits and dividends.  We
exclude Norway from the study because the Oslo B{\o}rs is not part of the Nasdaq Nordic system
and its data are not available through the same public sources; following its
acquisition by Euronext in 2019, data redistribution is subject to a distinct
licensing framework that restricts free access to historical and delayed
prices. Consequently, Oslo-specific indices (e.g., OSEBX, OSEAX) are not
covered by Google Finance and are therefore excluded from the sample.

The raw data are cleaned by removing stocks with extended information gaps. 
Short gaps are filled forward to maintain continuity. At time $t$, the vector of 
log-returns is computed as
\begin{equation}
\label{eqn:log_returns}
    \br(t)=\ln \bp(t)-\ln \bp(t-1),
\end{equation}
where $\bp(t)$ denotes the vector of closing prices. Summary
statistics of the resulting log-returns are reported in Table~\ref{tab:summary_data}.  
Consistent with established financial stylized facts, the series exhibit high 
kurtosis and negative skewness, indicative of heavy tails, volatility clustering 
and a non-Gaussian return distribution. These features have important implications 
for correlation estimation and the interpretation of eigenvalue dynamics, as will be 
seen later in Section~\ref{sec:results}.

\begin{table}
    \centering
    \caption{Considered stock markets and descriptive statistics of their log-returns.}
    \label{tab:summary_data}
    \begin{tabular}{ccccccccc}
        \hline
    \textbf{Index}  & $\boldsymbol{N}$   & \textbf{Days} &    \textbf{Mean}     & \textbf{Std. dev.} &     \textbf{Skewness}  &  \textbf{Kurtosis}   &   \textbf{Min}    &     \textbf{Max}  \\
        \hline
    {OMXS30} & 30       & 4,957     &  0.0003  &  0.0198  &  -0.2090   &  12.977    &  -0.2842      &  0.2768 \\
    {OMXC20} & 20       & 4,941     &  0.0003  &  0.0213  &  -0.1982   &  18.182    &   -0.3516     &  0.3771 \\
    {OMXH25} & 25       & 4,966     &  0.0001  &  0.0215  &  -0.2167   &  14.527    &   -0.4633     &   0.3498 \\
        \hline
    \end{tabular}
\end{table}

\subsection{Institutional Context}
The three indices considered herein are part of the Nasdaq Nordic equity marketplace,
which integrates Swedish, Danish, Finnish and Icelandic stock exchanges under a
common listing and trading platform. Although Nordic economies share
similar regulatory traditions and macroeconomic characteristics, they differ in
sector composition, financial depth, and sensitivity to external shocks. 
Together, the indices represent the most actively traded large-cap stocks in
their respective national markets and form widely used benchmarks for Nordic
equity exposure.

The OMXS30, launched in 1986, tracks the 30 largest and most traded securities
on Nasdaq Stockholm and is a modified free-float capitalization-weighted index,
reviewed semi-annually. The OMXC25, introduced in 2017 as a replacement for the 
OMXC20 benchmark, comprises the 25 most actively traded Danish equities listed on 
Nasdaq Copenhagen. However, to maintain a longer and consistent historical sample, 
I restrict attention to the OMXC20 subset.  Similarly, the OMXH25, established in 1988, 
includes the 25 most active Finnish stocks on Nasdaq Helsinki, is rebalanced quarterly, 
and applies a 10\% constituent cap. These indices serve as key indicators of national market
conditions, regional integration and exposure to global economic developments.

Evidence from Nordic macroeconomic history indicates that the region consists of
small open economies that are highly exposed to global financial turbulence, 
with stress episodes transmitting rapidly through trade, banking and commodity
channels. \citet{calmfors2024economic} discuss how Nordic economies have navigated 
major recent disturbances, in particular the Covid-19 collapse and the subsequent 
inflation and energy shock, against a broader backdrop that includes the Global Financial Crisis 
and the euro-area sovereign-debt episode. \citet{imf2013nordic} further 
emphasizes that the Nordic countries’ trade and financial openness implies substantial exposure 
to global and regional shocks, even though robust institutional frameworks provide important buffers. 
In financial markets, such stress episodes are typically associated with explosive 
co-movements and heightened systemic risk, motivating the study of time-varying correlation 
structure and a regime-aware approach to investment strategies in Nordic equity markets.

\section{Methodology}
\label{sec:method}

This section describes the methodological approaches used for the analysis and portfolio designs presented later on 
in Sections~\ref{sec:portfolio_construction} and~\ref{sec:results}. 

\subsection{Correlation matrix estimation}
\label{sec:rolling_window}
I start with the construction of correlation matrices from 
the daily logarithmic returns of stocks in each financial market under consideration. The computation is 
implemented in a rolling-window manner---see \citep{song2011evolution,adams2017correlations} for details.
For each market under consideration, the return series are partitioned into overlapping windows of length $T$, 
with a fixed step size $\Delta$. Now, let $N$ be the number of stocks in the index. At time $t$, 
for each window $\calW(t)=\figbrc{t-T, \ldots, t-1}$, I construct a $T\times N$ matrix of asset log-returns,
\begin{equation}
\label{eqn:return_matrix}
    \bR(t) = [\br(t-T), \ldots, \br(t-1)]^{\top},
\end{equation}
whose elements are given in~\eqref{eqn:log_returns}. From this matrix, I compute an $N \times N$ Pearson correlation matrix $\bC(t)$ as follows,
\begin{equation}
\label{eqn:corr_matrix}
    \bC(t) = \frac{1}{T-1} \bZ^{\T}(t)\bZ(t),
\end{equation}
where $\bZ(t) = z\figbrc{\bR(t)}$ is the standardized matrix of returns. Note that such a window construction is based only on past information available up to time $t-1$, which ensures that the analysis is free from look-ahead bias.

\subsection{Eigenvalue analysis}
For the sake of convenience, I drop the time index $t$ hereafter, unless stated otherwise. Within each window and for each market, I perform the \textit{eigendecomposition} of the correlation matrix in~\eqref{eqn:corr_matrix}, 
\begin{equation}
\label{eqn:eigendecomposition}
    \bC = \bV \bLambda \bV^{\T},
\end{equation}
where $\bLambda = \diag\brc{[\lambda_1, \lambda_2, \ldots, \lambda_N]}$ is a diagonal matrix containing sorted eigenvalues of $\bC$ (i.e., $\lambda_1 \geq \lambda_2 \geq \ldots \geq \lambda_N$) and $\bV = [\bv_1, \bv_2, \ldots, \bv_N]$ is a matrix whose columns are eigenvectors of $\bC$ reshuffled according to their corresponding eigenvalues.

Define the spectral density of eigenvalues of $\bC$ as
\begin{equation}
    \label{eqn:empirical_eigenvalue_distribution}
    f_{\bC}\brc{\lambda} = \frac{1}{N} \frac{\d n_{\bC}(\lambda)}{\d\lambda},
\end{equation}
where $n_{\bC}(\lambda)$ is the number of eigenvalues of $\bC$ less than $\lambda$.

RMT~\citep{bouchaud2003theory} provides a benchmark for interpreting this spectrum. 
In particular, under the null hypothesis that asset returns are uncorrelated,
the sample correlation matrix $\bC$ is asymptotically equivalent to a Wishart random
matrix. In the \emph{large-system limit}, where both the number of assets $N$ and the
sample size $T$ tend to infinity at a constant rate $c = N/T$, the empirical
eigenvalue density of $\bC$ converges to the MP distribution~\citep{marchenko1967distribution},
\begin{equation}
\label{eqn:mp_law}
    f\brc{\lambda} = \frac{\sqrt{\brc{\lambda_{+} - \lambda}\brc{\lambda-\lambda_{-}}}}{2\pi c\lambda}, \qquad \lambda\in[\lambda_{-},\lambda_{+}].
\end{equation}
This distribution represents the eigenvalue spread created purely by sampling noise when no true correlations exist. Thus, in empirical financial data, eigenvalues lying inside the MP interval $[\lambda_-,\lambda_+]$, where
\begin{equation}
\label{eqn:mp_bounds}
    \lambda_{\pm} = \left(1 \pm \sqrt{c}\right)^2,
\end{equation}
are regarded as compatible with estimation noise, whereas eigenvalues outside the MP bounds reflect a genuine correlation structure in the underlying system and are therefore treated as informative signals in the market. 
The upper bound $\lambda_{+}$ of the MP distribution can therefore serve as a threshold for distinguishing between noise and genuine market signal~\citep{laloux1999noise, plerou2002random}. Earlier studies reported that the fourth largest eigenvalue of the correlation matrix often does not exceed $\lambda_{+}$ in real stock markets~\citep[e.g.,][]{molero2025random}. In general, the leading two to three eigenvalues are often economically interpretable: the dominant eigenvalue is typically associated with the market-wide component, whereas the next eigenvalues may reflect sectoral or regional structure~\citep{plerou2002random}.

\subsection{Eigenportfolio betas}
\label{sec:eigenportfolio_betas}

To test whether the leading eigenmode of the return correlation matrix corresponds to a market factor in the sense of~\citet{sharpe1964capital} single-index model, I construct \emph{eigenportfolios} from the top $K$ eigenvectors of the correlation matrix $\bC$ and estimate their market exposures via ordinary least squares (OLS) regressions against a market proxy.

Let $r$ be the linear return of a portfolio $\bw$ at time $t$. Given the vector of log-returns in~\eqref{eqn:log_returns}, $r$ is computed as
\begin{equation}
\label{eqn:lin_return}
    r = [\exp(\br)-1]^{\top} \bw.
\end{equation}
For each market and the rolling window $\calW(t)$, after the eigendecomposition~\eqref{eqn:eigendecomposition}, I take the eigenvectors corresponding to the top $K$ sorted eigenvalues. Because the eigenportfolios are based on the correlation matrix $\bC$ (and not on the covariance matrix $\bSigma$) the weighting on a given asset must be set inversely proportional to its volatility~\citep[see][]{avellaneda2010statistical}, 
\begin{equation}
\label{eqn:eigenportfolio}
    \bw^{\textrm{eig}}_k = \frac{\bv_k}{\sigma_k}, \qquad k = \figbrc{1, \ldots,K}.
\end{equation}
I then normalize the above eigenportfolio weights so that they sum up to 1. 

Let $r^{\textrm{eig}}_k(t)$ denote the linear return of the eigenportfolio $\bw^{\textrm{eig}}_k$ computed using~\eqref{eqn:lin_return} and $r^{\textrm{mkt}}(t)$ denote the corresponding market return. A proxy for the latter is given by the return of the \textit{equal-weight} (EW) portfolio over the same constituents. The portfolio weights are thus given by
\begin{equation}
\label{eqn:ew_portfolio}
    w^{\textrm{EW}}_n = \frac{1}{N}, \qquad n = 1,\dots,N,
\end{equation}
while the market returns are calculated as
\begin{equation}
\label{eqn:market_return}
    r^{\textrm{mkt}} = r^{\textrm{EW}} = [\exp(\br)-1]^{\top} \bw^{\textrm{EW}}.
\end{equation}

Given the above, I estimate for each $k\in\{1,\ldots,K\}$ the univariate OLS regression
\begin{equation}
\label{eqn:method_market_factor_ols}
    r^{\textrm{mkt}}(t) = \alpha_k + \beta_k\, r^{\textrm{eig}}_k(t) + \varepsilon_k(t), \quad t = \figbrc{1,\ldots, M},
\end{equation}
where $M$ is the total number of weeks in the observation period, while $\alpha_k$ and $\beta_k$ are regression coefficients and $\varepsilon_k(t)$ is an i.i.d. disturbance with zero mean. The slope $\beta_k$ can equivalently be expressed as
\begin{equation}
\label{eqn:beta}
    \beta_k = \frac{\mathrm{cov}\figbrc{r^{\textrm{eig}}_k,r^{\textrm{mkt}}}}{\mathrm{var}\figbrc{r^{\textrm{mkt}}_k}}.
\end{equation}
In analogy with the classical result that the market portfolio has unit loading on the market factor, an eigenportfolio consistent with the market mode is expected to exhibit $\beta_k\approx 1$ (up to the chosen eigenportfolio normalization) and high explanatory power for $r^{\textrm{mkt}}(t)$. 

\subsection{Correlation matrix cleaning}
Empirical correlation and covariance matrices estimated from historical returns are notorious for containing substantial estimation noise, especially in high-dimensional settings where the number of assets is large relative to
the sample length~\citep[see][Ch.2]{deprado2020machine}. This noise leads to numerical ill-conditioning and adversely affects downstream tasks such as portfolio optimization. 

A common remedy for numerical instability in empirical correlation estimates builds the observation of \citet{laloux1999noise} that the noise component of the sample correlation matrix $\bC$ is well approximated by the MP law~\eqref{eqn:mp_law}. Under this interpretation, only a small set of eigenvalues exceeding the MP upper edge is deemed informative and associated with common risk factors. Building on this insight, \citet{laloux2000random} propose a \textit{denoising} procedure in which the eigenvalues within the MP bulk are replaced by a single constant, chosen to preserve the trace of $\bC$, while the outlying eigenvalues above the MP bound are left unchanged. In practice, this spectral clipping improves numerical stability by mitigating ill-conditioning and increases statistical efficiency by attenuating noise-dominated modes, yielding more reliable dependence estimates and improved out-of-sample performance in risk measurement and portfolio construction \citep{pafka2003noisy}.

Concretely, let $k^{\star}$ denote the index of the last eigenvalue above the MP upper bound, i.e., $\lambda_{k^{\star}} > \lambda_{+}$. The cleaned set of eigenvalues is constructed by retaining $\lambda_1,\ldots,\lambda_{k^{\star}}$ and replacing the set of remaining eigenvalues by it average,
\begin{equation}
\label{eqn:lambdas_clipped}
    \tilde{\lambda}_k=
    \begin{cases}
    \lambda_k, & k \le k^{\star},\\
    \dfrac{1}{N-k^{\star}}\displaystyle\sum\limits_{j=k^{\star}+1}^{N}\lambda_j, & k>k^{\star},
    \end{cases}
\end{equation}
thus preserving the trace of $\bC$.  Given the matrix of eigenvectors $\bV$, the preliminary denoised matrix is
\begin{equation}
    \tilde{\bC} = \bV\tilde{\bLambda}\bV^\top,    
\end{equation}
where $\tilde{\bLambda}$ is the diagonal matrix of the cleaned eigenvalues in~\eqref{eqn:lambdas_clipped}. Because this transformation perturbs the diagonal elements of $\bC$, a final rescaling step is applied to restore unit variance,
\begin{equation}
\label{eqn:denoised_corr_mat}
    \bC_{\mathrm{den}}
=
\bS^{-1/2}
\,\tilde{\bC}\,
\bS^{-1/2},
\end{equation}
where $\bS = \operatorname{diag}(\tilde{\bC})$. In this way, the obtained $\bC_{\mathrm{den}}$ is again a proper correlation matrix and can be readily used for computing portfolio allocations. 

A sample covariance matrix is obtained from a correlation matrix $\bC$ as follows,
\begin{equation}
\label{eqn:cov_corr}
    \bSigma=\bD^{\top}\bC\bD,
\end{equation}
where $\bD = \diag\brc{[\sigma_1, \ldots, \sigma_N]}$ is a diagonal matrix of asset volatilities. 
Covariance matrix $\bSigma$ will be used later in Section~\ref{sec:baselines}. 
An additional step in cleaning it is the application of \emph{shrinkage} \citep{ledoit2004well} towards a target matrix $\bT$. The target may be chosen to reflect a parsimonious prior belief about the covariance structure in the market under consideration (e.g., identity or equicovariance matrix). For instance, one may perform
\begin{equation}
\label{eqn:shrinkage}
    \tilde{\bSigma} = (1-\delta) \bSigma + \delta \bT,
\end{equation}
where $\delta\in\sqbrc{0,1}$ is the shrinkage intensity, while $\bT$ may be chosen as a compound-symmetry target, given by
\begin{equation}
\label{eqn:const_cov_target}
    \sqbrc{\bT}_{i,j}=
\begin{cases}
\frac{1}{N}\sum_{i=1}^{N}\sqbrc{\bSigma}_{i,i}, & i=j,\\
\frac{2}{N(N-1)}\sum_{1\le i'<j'\le N}\sqbrc{\bSigma}_{i',j'},     & i\neq j.
\end{cases}
\end{equation}
That is, all diagonal elements of $\bT$ equal the average variance, while all off-diagonal elements equal the average covariance. This parsimonious structure reduces estimation noise by shrinking the sample covariance matrix toward a homogeneous covariance specification, with the Ledoit–Wolf estimator applying the shrinkage intensity to trade off bias and variance~\citep{ledoit2004well}.

\subsection{Portfolio performance assessment}
To assess the performance of portfolios under consideration, the aforementioned rolling-window experiment is performed. That is, for each window $\calW(t)$, I compute the correlation matrix $\bC$ of log-returns using~\eqref{eqn:corr_matrix}. Based on $\bC$, I compute the corresponding portfolio weight vector $\bw$ for each strategy under consideration. To assess the return developments, I compute the series of cumulative returns of each portfolio as follows,
\begin{equation}
    R(t) = \prod_{\tau=1}^{t} \sqbrc{1 + r(\tau)} - 1,
\end{equation}
where the linear return at time $t$ is computed using~\eqref{eqn:lin_return}.

Moreover, the performance is also reported by a series of additional metrics. For instance, assuming---without loss of generality---zero risk-free rate and frictionless weekly rebalancing, annualized mean return, volatility and Sharpe ratio of a portfolio are computed as follows,
\begin{align}
    \mu &= \brc{ \prod_{\tau=1}^{M} \sqbrc{1 + r(\tau)} }^{\frac{52}{M}} - 1, \\
    \sigma &= \sqrt{\frac{52}{M-1} \sum_{\tau=1}^{M} \sqbrc{r(\tau) - \mu}^2}, \\
    \textrm{Sharpe} &= \mu / \sigma,
\end{align}
where $M$ is the total number of weeks in the observation period, and 52 is the number of weeks in a year.

In addition to the above, the Sortino ratio is defined as a downside-risk–adjusted performance measure that penalizes only negative return deviations, thereby addressing a key limitation of the Sharpe ratio, which treats upside and downside volatility symmetrically. The Sortino ratio is computed as
\begin{equation}
\mathrm{Sortino}
=
\mu / \sqrt{\frac{52}{M-1} \sum_{\tau=1}^{M} \sqbrc{\min\figbrc{r(\tau), 0}}^2}.
\end{equation}
The denominator corresponds to the downside deviation, computed using only returns below the target threshold (zero in this case). By construction, the Sortino ratio focuses on adverse outcomes and is therefore particularly relevant for evaluating strategies designed to mitigate downside risk.

The Treynor ratio measures excess return per unit of systematic risk, where systematic risk is quantified by the portfolio's beta with respect to a chosen market proxy. Let $r^{\textrm{mkt}}$ denote the market return computed from the market EW proxy using~\eqref{eqn:lin_return}. The portfolio beta is then estimated using~\eqref{eqn:beta} and the Treynor ratio is then defined as
\begin{equation}
\mathrm{Treynor}
=
\mu / \beta.
\end{equation}
Under this definition, the Treynor ratio of the EW portfolio coincides with its return by construction (since $\beta^{\textrm{EW}}=1$) and therefore serves as a natural reference level. For alternative strategies, the Treynor ratio quantifies the efficiency with which returns are generated per unit of exposure to the common component of the market.

\subsection{Baseline portfolios}
\label{sec:baselines}
The following portfolios are considered as baselines in this study.

The EW portfolio assigns identical weights to all assets in the investment universe, regardless of their volatility, size or any other characteristics. The EW portfolio weights are computed using~\eqref{eqn:ew_portfolio}. As one of the simplest diversification schemes, this portfolio is widely used as a benchmark in empirical asset-pricing and portfolio-construction studies, and it often performs competitively due to its implicit contrarian rebalancing mechanism~\citep{demiguel2009optimal}. Despite its simplicity, the EW allocation provides a robust baseline for evaluating more sophisticated portfolio-design methods.

A further baseline considered in this study is the principal eigenportfolio, constructed from the eigenvector associated with the leading eigenvalue $\lambda_1$ of the correlation matrix $\bC$. This dominant component is commonly interpreted as the market mode, capturing broad market-wide co-movement and typically accounting for a substantial share of cross-sectional variation in tranquil regimes. The corresponding portfolio weights are computed according to~\eqref{eqn:eigenportfolio} and normalized so that they sum up to 1. As noted by~\citet{avellaneda2010statistical}, since the principal eigenportfolio weights are typically positive and scale inversely with asset volatility, such pattern is broadly consistent with capitalization-weighted allocations insofar as larger firms tend to exhibit lower return volatility. 

The weights of the above two baseline schemes lie in the \textit{probability simplex} $\calP$,
\begin{equation}
\label{eqn:prob_simplex}
    \calP = \figbrc{\,\bw \ge \bzero,\; \bw^{\top} \bone = 1 \,}.
\end{equation}
Thus, the allocations are long-only and satisfy the full-investment (budget) constraint, ensuring that the resulting portfolios are directly implementable and investable, which enables transparent comparisons with market indices and alternative portfolio constructions.

The \textit{minimum-variance} (MV) portfolio is one of the foundational constructs in modern portfolio theory. Introduced by \citet{markowitz1952portfolio} as part of his mean-variance paradigm, it represents the portfolio with the lowest possible return variance among all fully invested portfolios. Formally, the MV portfolio solves an optimization problem under a full-investment constraint,
\begin{equation}
\label{eqn:mv_portfolio}
    \bw^{\textrm{MV}} = 
    \arg\min_{\bw \in \calP} \figbrc{\bw^\top \bSigma \bw},
\end{equation}
where $\calP$ is the probability simplex~\eqref{eqn:prob_simplex}. Practically, this problem can be solved by convex optimization toolboxes~\citep[see, e.g.,][]{grant2009cvx, mathworks2025quadprog}. 

The MV portfolio is attractive for both theoretical and practical reasons. Because it does not rely on estimates of expected returns, its construction is substantially more stable and robust than Markowitz's original mean-variance portfolio, which are notoriously sensitive to estimation error~\citep{merton1980estimating}. By relying only on the covariance matrix $\bSigma$, the MV portfolio tends to tilt toward low-volatility and weakly correlated assets, thereby generating diversification benefits even when return forecasts are unreliable. Empirically, a large body of literature shows that minimum-variance portfolios often outperform na\"{i}ve diversification and many more complex strategies on a risk-adjusted basis \citep[see, e.g.,][]{jagannathan2003risk,clarke2006minimum}.

\section{Regime-aware portfolio construction}
\label{sec:portfolio_construction}
In this section, I present the proposed approach to portfolio optimization, taking into account the market regime. Inspired by the findings of~\cite{molero2025random}, where the idea of a ``safe haven'' for times of high volatility was presented, I aim at constructing a defensive portfolio that would provide a means of diversification for such stressful periods. 

\subsection{Market crisis detection}
\label{sec:crisis_detection}
Market crises are commonly associated with surges in cross-asset co-movement, driven by contagion, herding and liquidity shocks propagating through globally integrated markets~\citep{bekaert2003market, chiang2007dynamic}. Even in the absence of a domestic crisis, externally induced uncertainty---such as geopolitical events or global risk-off episodes---can cause assets to move in an unusually synchronized manner. During such episodes, systematic risk factors dominate the idiosyncratic variation, institutional investors rebalance portfolios in similar ways and correlations rise above levels justified by economic fundamentals. This erosion of cross-sectional heterogeneity reduces the effectiveness of diversification. Nordic markets, being an export-oriented and internationally exposed equity universe, are particularly susceptible to these externally driven correlation spikes~\citep{fjaervik2023crash}.

As mentioned earlier, a well-established empirical regularity in the RMT literature is that market-wide stress manifests itself as a sharp increase in the leading eigenvalue of the correlation matrix~\citep{laloux1999noise, plerou2002random}. Building on this insight, I introduce a regime-classification measure based on the relative dominance of the leading eigenvalue, the \textit{eigenvalue ratio}, defined as
\begin{equation}
\label{eqn:crisis_indicator}
    \chi(t) = z\figbrc{\lambda_{1}(t) / \lambda_{2}(t)},
\end{equation}
which contrasts the market mode with the strongest counter-movement mode. Large values of $\chi(t)$ indicate heightened synchronization and thus crisis-like conditions, whereas smaller values reflect more diversified and calm regimes. 

To obtain a practical and stable classification signal, I detrend $\chi(t)$, smoothening it using a $\ell$-day moving average and threshold it over zero. Specifically, the system is classified as being in a \textit{crisis} regime whenever the smoothed and detrended ratio~\eqref{eqn:crisis_indicator} is non-negative and in a \textit{calm} regime otherwise. This procedure yields a transparent, data-driven segmentation of market dynamics based solely on the competition between dominant eigenmodes in the correlation structure.

\subsection{Regime-aware strategy}
The behavior of cross--asset correlations depends strongly on the prevailing
market regime.  In calm periods, correlations are moderate and the empirical
covariance matrix is typically well conditioned.  In crisis episodes, however,
correlations surge and a single market mode can dominate the correlation
structure, leading to near singularity and unstable minimum--variance
allocations.  To accommodate these features, I construct portfolios in an
adaptive manner according to the following steps:

\begin{enumerate}
    \item \textbf{Regime identification.} The aforementioned market regime 
    indicator~\eqref{eqn:crisis_indicator} classifies each observation window 
    ($\chi(t)\gtrless0$) as either calm or crisis. The relative dominance 
    of the first eigenvalue over the second eigenvalue provides a stable proxy for 
    market-wide co-movement.

    \item \textbf{Covariance cleaning.} The empirical correlation matrix $\bC$ is denoised 
    via~\eqref{eqn:lambdas_clipped},  which suppress sampling noise in the eigenvalue spectrum 
    while preserving economically relevant dependence. The covariance matrix $\bSigma$
    is then computed using~\eqref{eqn:cov_corr} and shrunk towards equicovariance using Ledoit-Wolf 
    shrinkage~\eqref{eqn:shrinkage}.

    \item \textbf{Regime-aware portfolio design.}
    \begin{enumerate}
        \item \emph{Calm periods.} The obtained clean covariance matrix is then
        used to construct a capable standard portfolio, e.g., MV allocation~\eqref{eqn:mv_portfolio}.
        \item \emph{Crisis periods.} A natural implication of the increased market-wide co-movement 
        is to cap portfolio exposure to the dominant (market) eigenmode 
        in stressed regimes. Moreover, the evidence of \citet{molero2025random} that the second 
        eigenmode can exhibit defensive behavior in high-volatility periods motivates 
        maintaining a minimum allocation to the subleading eigenmode as a defensive component. 
        Operationally, this can be implemented by complementing the probability-simplex constraint 
        in the portfolio optimization problem with additional constraints on portfolio exposures to 
        the leading and subleading eigenmodes. For the case of MV allocation, the portfolio 
        optimization problem can be re-formulated as follows,
        \begin{equation}
        \label{eqn:regime_aware_problem}
            \begin{aligned}
                & \underset{\bw\in\real{N}}{\textrm{minimize}} & \bw^{\top} \bSigma \bw\\
                & \textrm{subject to} & \bw^{\top} \bv_1 \leq \gamma_1 \\
                & & \bw^{\top} \bv_2 \geq \gamma_2 \\
                & & \bw \in \calP.
            \end{aligned}
        \end{equation}
        This objective yields a quadratic program, which can be solved using standard convex 
        optimization toolboxes~\citep[see, e.g.,][]{grant2009cvx, mathworks2025quadprog}.
    \end{enumerate}
\end{enumerate}

The proposed regime-aware strategy is designed to preserve the advantages of a conventional portfolio allocation 
(in this case Markowitz MV portfolio) in normal conditions while mitigating its vulnerability to factor dominance 
and exploiting defensive structures during stressed periods. By further combining denoising and shrinkage, 
the procedure yields more stable correlation and covariance estimates and more robust allocations under 
varying market conditions.

\section{Results and discussion}
\label{sec:results}
This section presents the results of the study based on the rolling-window experiment described in Section~\ref{sec:method}. For the analysis, I choose the window size to be $T=252$ days (i.e., one year of trading days), while the window step size to be $\Delta=5$ days (i.e., one trading week). The correlation matrix $\bC$ of log-returns is computed according to the steps presented in Section~\ref{sec:rolling_window}, the eigenvalues are computed and utilized for the analysis, regime detection and portfolio design.

\subsection{Spectral dynamics}
\begin{figure*}[t]
	\centering
	\subcaptionbox{OMX Stockholm 30.\label{fig:plot_eig_omxs30}}[0.35\linewidth]
	{\centering
		{\includegraphics[height=3.6cm, trim={0.4cm 0cm 1cm 0cm}, clip=true]{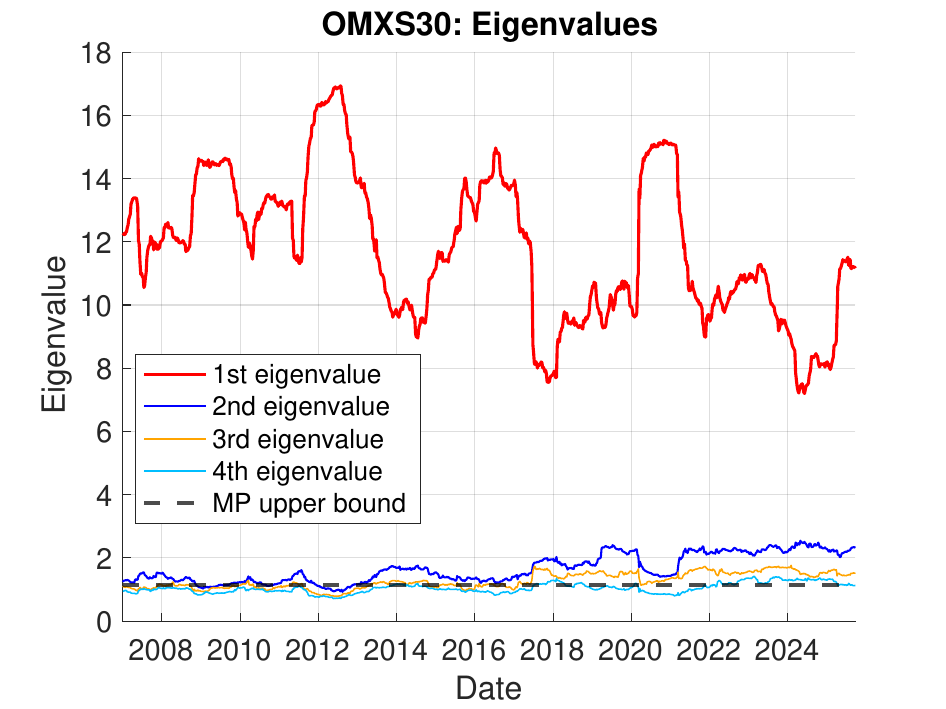}}
		}%
	\subcaptionbox{OMX Copenhagen 20.\label{fig:plot_eig_omxc20}}[0.31\linewidth]
	{\centering
		{\includegraphics[height=3.6cm, trim={1.55cm 0cm 1.65cm 0cm}, clip=true]{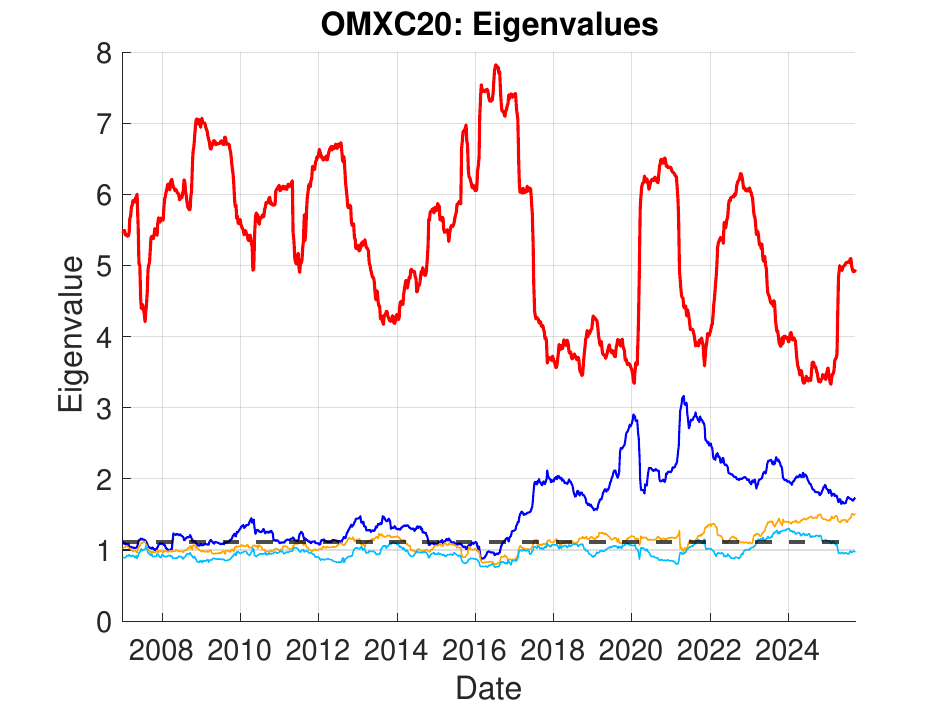}}
		}%
    \subcaptionbox{OMX Helsinki 25.\label{fig:plot_std_omxh25}}[0.33\linewidth]
	{\centering
		{\includegraphics[height=3.6cm, trim={1.3cm 0cm 1cm 0cm}, clip=true]{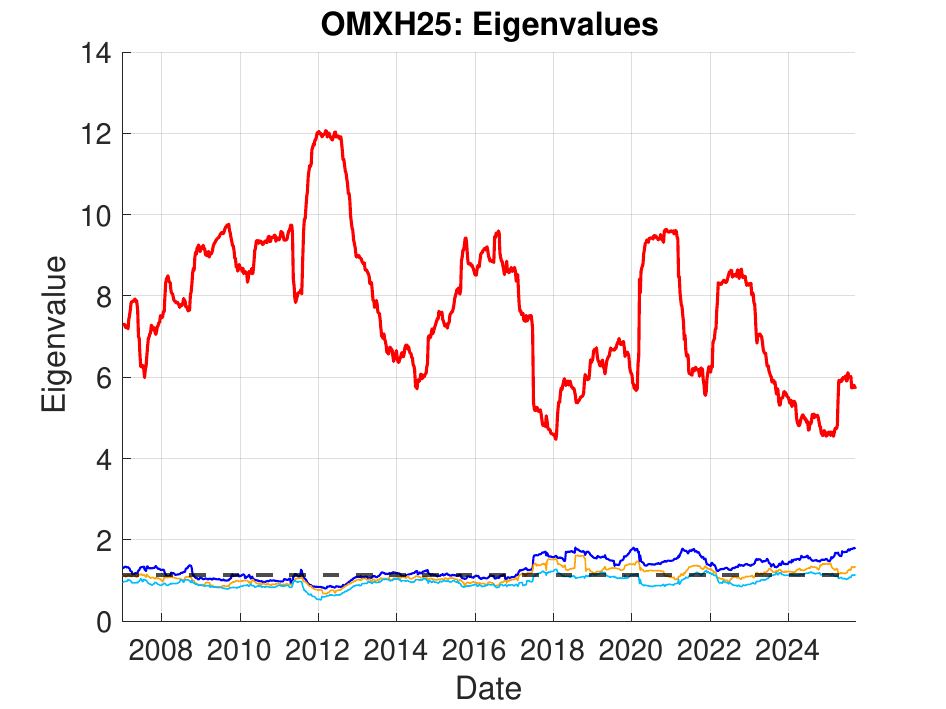}}
		}
	\caption{Time evolution of the four largest eigenvalues of the correlation matrix of log-returns of the Nordic markets.}
	\label{fig:dynamics_eigenvals}
\end{figure*}

I start with plotting the time evolution of the first (red), second (blue), third (orange) and fourth (cyan) largest eigenvalues of the correlation matrix in Figure~\ref{fig:dynamics_eigenvals}, alongside the MP upper bound $\lambda_{+}$ computed in~\eqref{eqn:mp_bounds}. The largest eigenvalue displays pronounced spikes at specific intervals, consistent with episodes of elevated market-wide co-movement and stronger cross-asset dependence. Notably, prior to ca. 2017, the second and third largest eigenvalues frequently lay below the MP upper bound across all markets considered. Also, more generally, it can be seen that these eigenvalues fluctuate persistently near the MP upper bound for the OMXS30 and OMXH25 indices, while there have been significant raises in the second eigenvalue in the OMXC20 index. Meanwhile, the fourth eigenvalue lies mostly below the MP upper bound, which is consistent with the observations of~\citet{molero2025random} for various stock markets. 

\begin{figure*}[t]
	\centering
	\subcaptionbox{OMX Stockholm 30.\label{fig:plot_std_eig_omxs30}}[0.35\linewidth]
	{\centering
		{\includegraphics[height=3.6cm, trim={0.4cm 0cm 1cm 0cm}, clip=true]{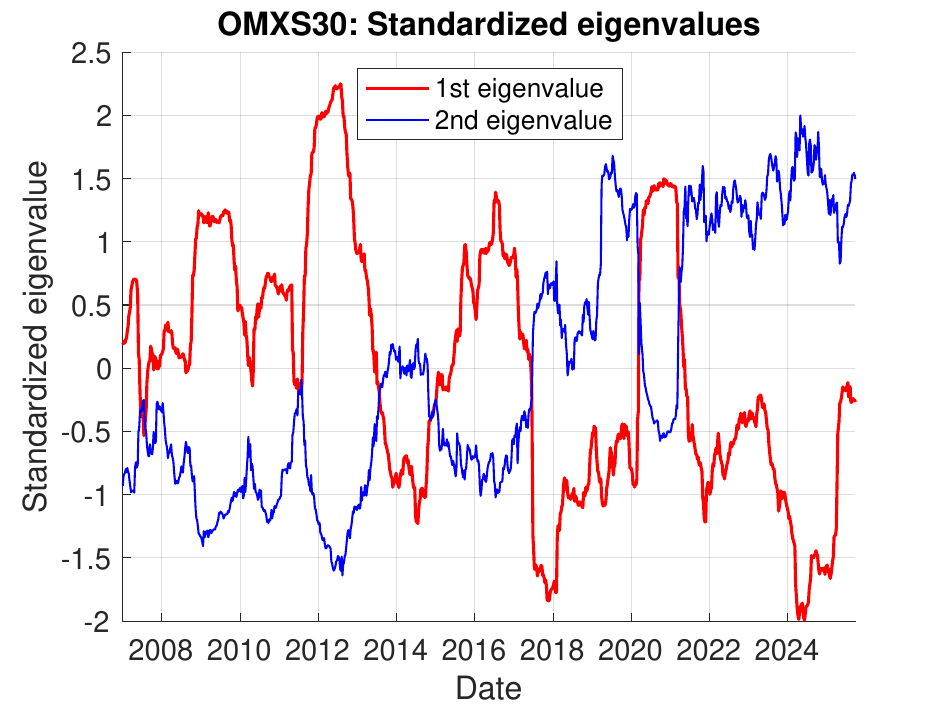}}
		}%
	\subcaptionbox{OMX Copenhagen 20.\label{fig:plot_std_eig_omxc20}}[0.31\linewidth]
	{\centering
		{\includegraphics[height=3.6cm, trim={1.55cm 0cm 1.65cm 0cm}, clip=true]{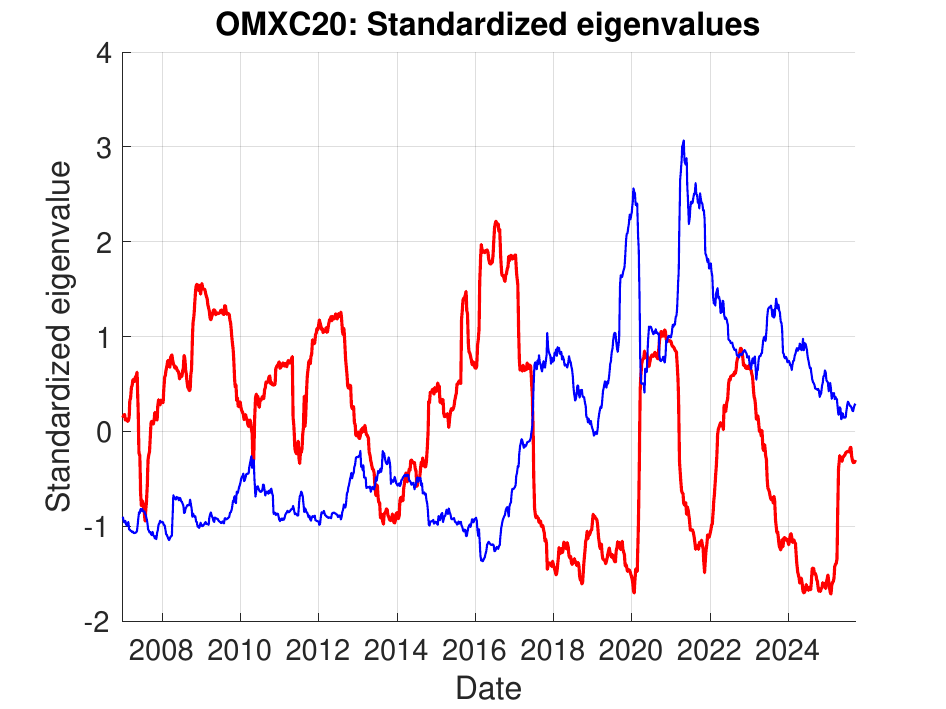}}
		}%
    \subcaptionbox{OMX Helsinki 25.\label{fig:plot_std_eig_omxh25}}[0.33\linewidth]
	{\centering
		{\includegraphics[height=3.6cm, trim={1.3cm 0cm 1cm 0cm}, clip=true]{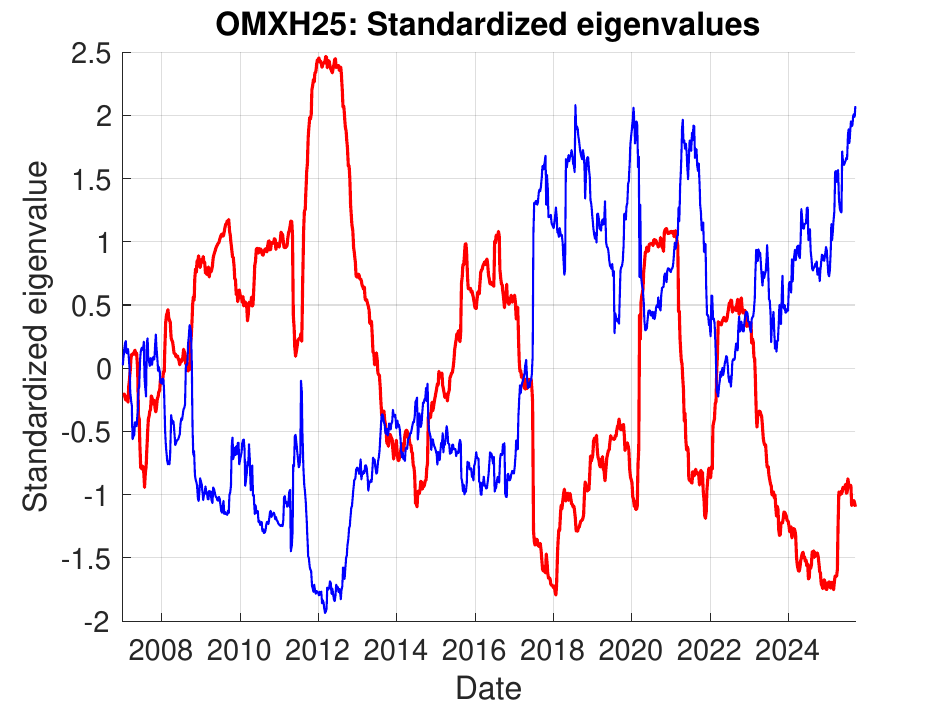}}
		}
 	\caption{Dynamics of the two largest eigenvalues of the log-return correlation matrix, standardized for comparison sake.}
	\label{fig:dynamics_std_eigenvals}
\end{figure*}

To dive deeper, I analyze the temporal dynamics of the standardized versions of the two largest eigenvalues of the correlation matrix of log-returns. Figure~\ref{fig:dynamics_std_eigenvals} displays their evolution, allowing direct comparison of their fluctuations on a common scale. A clear pattern is seen from the figure: the two largest eigenvalues exhibit a pronounced counter-phase relationship across all three considered markets. Therefore, I confirm the findings of~\citet{molero2025random} also for the case of Nordic stock markets. This systematic opposition indicates that the second eigenvalue, together with its associated eigenvector, can be thought of as a counter-risk component, potentially offering diversification benefits and serving as a hedge or divestment channel during periods of heightened market stress.

\begin{figure}[t!]
    \centering
    \begin{minipage}[t]{0.48\textwidth}
        \centering
        \includegraphics[width=\linewidth]{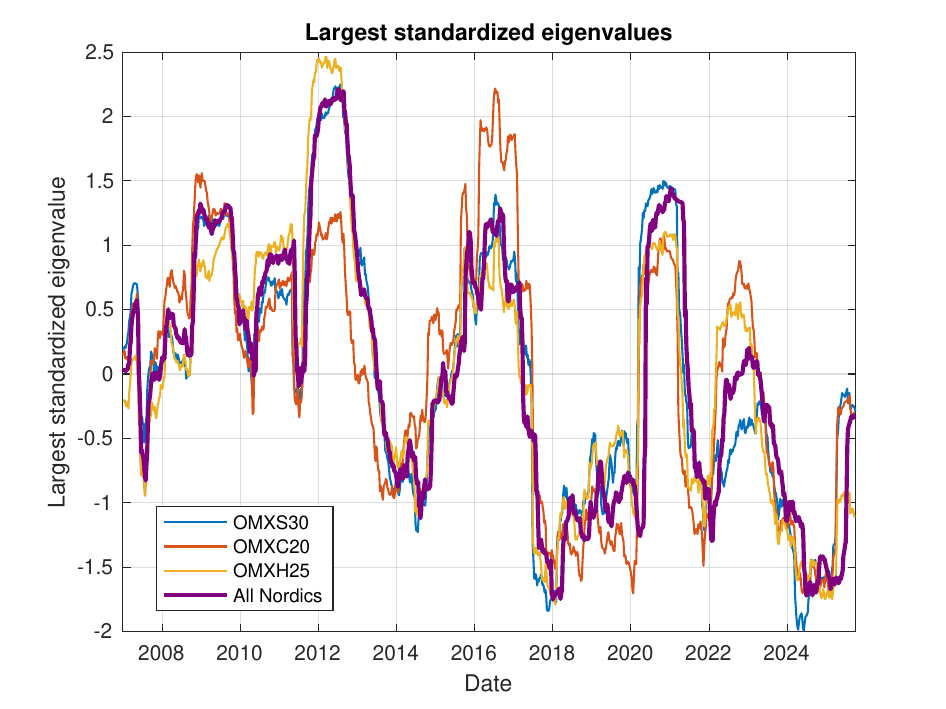}
        \caption{Dynamics of the largest standardized eigenvalue of the correlation matrix of the Nordic markets.}
        \label{fig:plot_comparison_nordics}
    \end{minipage}
    \hfill
    \begin{minipage}[t]{0.48\textwidth}
        \centering
        \includegraphics[width=\linewidth]{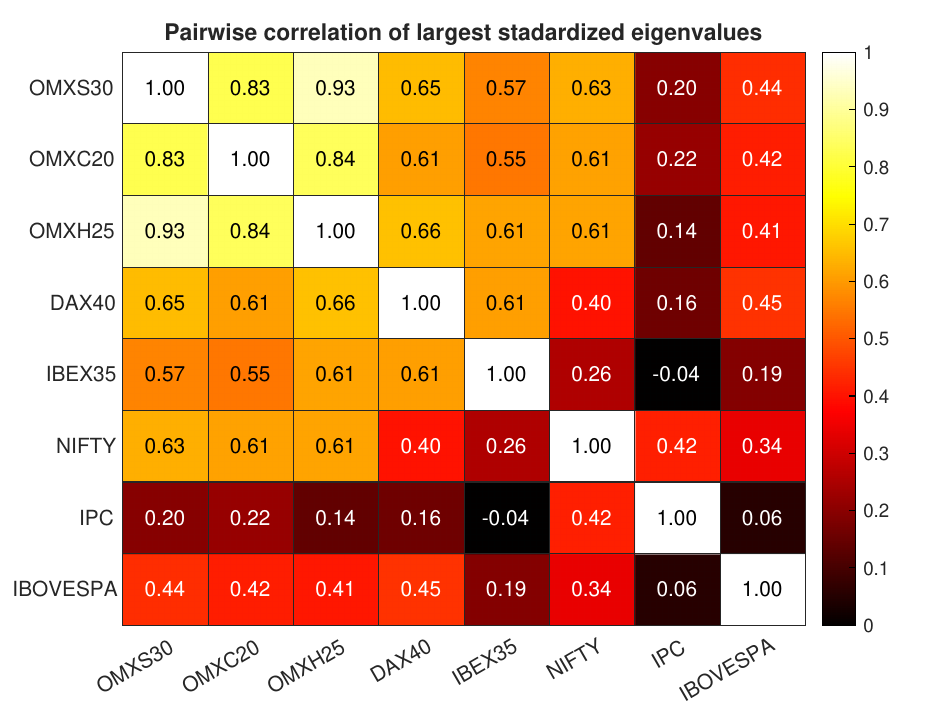} 
        \caption{Cross-correlation among the largest standardized eigenvalues of the correlation matrices of diverse markets.}
        \label{fig:plot_corr_world_markets}
    \end{minipage}
\end{figure}

Building on these observations, Figure~\ref{fig:plot_comparison_nordics} shows the temporal evolution of the largest standardized eigenvalue for the three Nordic markets, alongside an \textit{aggregated Nordic index} constructed by combining all tickers from OMXS30, OMXC20 and OMXH25. Note that in order to construct the aggregated index, I retime and aggregate the data of individual markets on a weekly basis to synchronize the corresponding time series. 

It can be seen from the figure that the three individual markets exhibit a strikingly synchronized behavior: their eigenvalue peaks (crisis periods) and troughs (calm periods) occur almost simultaneously, indicating that systemic risk in the region might be driven by shared regional or global shocks rather than by country-specific dynamics. The aggregated Nordic market (purple curve) closely follows this shared behavior while exhibiting reduced idiosyncratic noise, effectively providing a cleaner representation of the underlying regional market behavior. Notably, the Swedish OMXS30 follows the aggregate series most closely, which might be due to its dominant share of market capitalization and liquidity in the region. These findings suggest that the Nordic markets form a tightly coupled system with highly correlated correlation dynamics. 

Motivated by this observation, I compute pairwise correlations between the standardized largest eigenvalues of the correlation matrices of the three Nordic markets, as well as those of a broader set of international markets, including DAX40 (Germany), IBEX35 (Spain), NIFTY (India), IPC (Mexico) and IBOVESPA (Brazil). The resulting correlation structure, shown in Figure~\ref{fig:plot_corr_world_markets}, reveals a distinct cluster of strong correlations among the Nordic markets, consistent with geographic proximity as well as dense economic linkages and trade integration. The figure also shows that European markets per se also exhibit relatively high mutual correlations, indicating that systemic risk pressures tend to propagate more uniformly within the region. Meanwhile, correlations between European and non-European markets are noticeably weaker and more heterogeneous, suggesting that the dominant eigenvalue captures regional rather than fully global systemic dynamics. Emerging markets, such as those of India, Mexico and Brazil, exhibit markedly lower co-movement with the Nordic and broader European indices, as well as with one another, which is consistent with more heterogeneous and region-specific risk exposures.

\subsection{Market factor analysis}
To assess whether the dominant eigenmode of the return correlation matrix admits a market-factor interpretation, I follow the eigenportfolio-based approach described in Section~\ref{sec:eigenportfolio_betas}. Specifically, I examine the linear relationship between the market return and the returns of $K=3$ eigenportfolios constructed from the corresponding leading eigenvectors. For each $k=1,2,3$, I estimate the univariate OLS regression in~\eqref{eqn:method_market_factor_ols} and report $\beta_k$, the coefficient of determination $R^2$, and the p-value for testing the null-hypothesis $H_0:\beta_k=0$ in Table~\ref{tab:nordic_eigenbeta}. These statistics jointly quantify the magnitude and statistical significance of the market-factor interpretation of the leading eigenmodes.

\begin{table}[t!]
\centering
\renewcommand{\arraystretch}{1.15}
\setlength{\tabcolsep}{6pt}
\caption{OLS regressions of market returns on eigenportfolio returns.}
\label{tab:nordic_eigenbeta}
\begin{tabular}{l rrr rrr rrr}
\hline
& \multicolumn{3}{c}{\textbf{OMXS30}} &
  \multicolumn{3}{c}{\textbf{OMXC20}} &
  \multicolumn{3}{c}{\textbf{OMXH25}} \\
\textbf{} &
{$\beta$} & {$R^2$} & {p-val} &
{$\beta$} & {$R^2$} & {p-val} &
{$\beta$} & {$R^2$} & {p-val} \\
\hline
$\lambda_1$ & \textbf{0.936} & 0.980 & \multicolumn{1}{r}{$<0.001$} &
    \textbf{0.972} & 0.944 & \multicolumn{1}{r}{$<0.001$} &
    \textbf{0.984} & 0.984 & \multicolumn{1}{r}{$<0.001$} \\
$\lambda_2$ & 0.155 & 0.008 & 0.166 &
    0.162 & 0.010 & 0.114 &
    0.043 & 0.001 & 0.680 \\
$\lambda_3$ & 0.062 & 0.001 & 0.670 &
    0.293 & 0.030 & 0.006 &
   -0.041 & 0.001 & 0.695 \\
\hline
\end{tabular}
Note: Betas of the principal eigenportfolio are highlighted in bold font.
\end{table}

In the classical single-factor framework~\citep{sharpe1964capital}, if the first eigenportfolio spans the market mode, its return should closely track the market return, implying a slope near unity and a high coefficient of determination. Table~\ref{tab:nordic_eigenbeta} shows that for OMXS30, OMXC20 and OMXH25, the principal eigenportfolio exhibits \(\beta_1\in[0.936,0.984]\) and explains most of the variation in the market return (\(R^2\in[0.944,0.984]\)), with coefficients that are highly statistically significant. This combination of near-unit loading and strong explanatory power supports interpreting the leading eigenvector as a market mode, i.e., a pervasive common component shared by the majority of constituents.

In contrast, the second eigenportfolio carries only a small and statistically insignificant loading on the market return in all three markets, with negligible explanatory power ($R^2 \simeq 10^{-3}$--$10^{-2}$). This indicates that the second eigenmode is largely orthogonal to the market factor and does not contribute materially to the co-movement captured by the market portfolio. The third eigenportfolio is likewise weak in economic terms: for OMXS30 and OMXH25 it is essentially unrelated to the market return, while for OMXC20 it yields a statistically significant but small loading ($\beta_3 \approx 0.29$) and a modest $R^2 \approx 0.03$. Overall, the evidence points to a pronounced market mode in the Nordic equities considered, with the first eigenmode accounting for most market-wide co-movement and largely spanning the aggregate market return, while higher-order eigenmodes contribute less to market-return variation but may still carry economically relevant information—particularly for defensive positioning in stress regimes

\subsection{Crisis indicator performance}
\begin{figure}[t]
	\centering
		\includegraphics[width=0.5\linewidth]{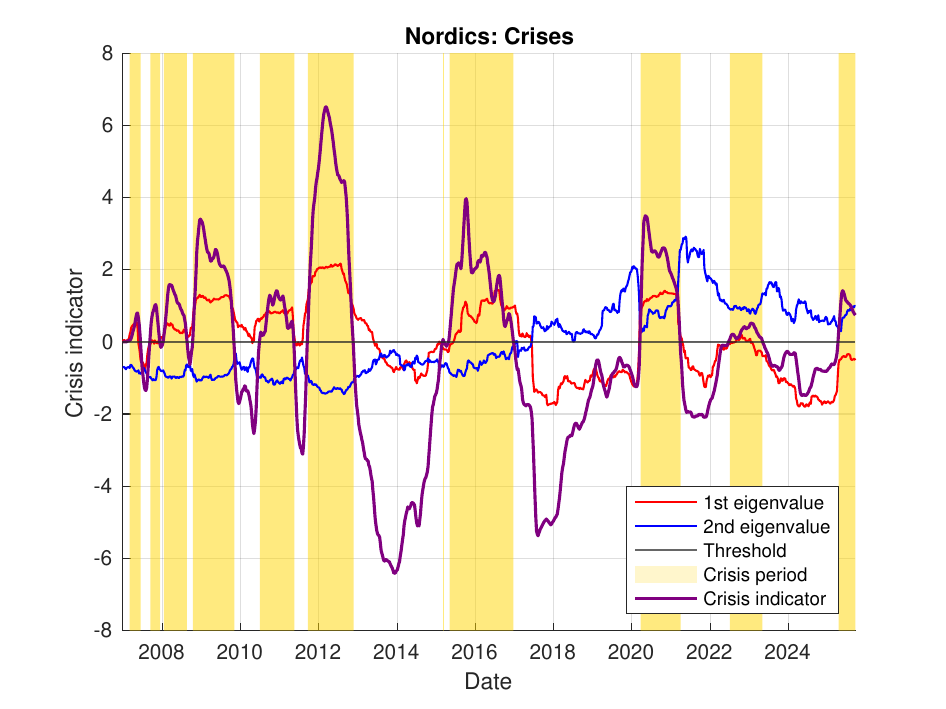}
        \caption{Market regime classification using the crisis indicator $\chi(t)$ and its relation to the dynamics of the standardized leading two eigenvalues for the aggregated Nordic stock market.}
	\label{fig:plot_crisis_indicator}
\end{figure}

In Section~\ref{sec:crisis_detection}, I have introduced the crisis indicator in~\eqref{eqn:crisis_indicator}, which discriminates between two market regimes. To illustrate its behavior, Figure~\ref{fig:plot_crisis_indicator} plots the eigenvalue ratio $\chi(t)$ (purple) alongside the two standardized leading eigenvalues $\lambda_1(t)$ and $\lambda_2(t)$ (red and blue, respectively) of the aforementioned aggregated Nordic market correlation matrix. The series $\chi(t)$ is smoothed using a moving-average filter with window length $\ell=10$ trading days. Regimes are then identified by thresholding $\chi(t)$ at zero (black horizontal line), classifying observations with $\chi(t)\ge 0$ as crisis periods (yellow shading) and $\chi(t)<0$ as calm periods (white background).

By examining the figure, one firstly sees that periods classified as crises coincide with rises of the leading eigenvalue and falls of second eigenvalue. Moreover, one can see a connection of these crises to several actual episodes of stress experienced by the Nordic stock markets, closely tied to global crises. In 2008--2009, the Global Financial Crisis triggered sharp declines across all Nordic indices, with huge losses and pronounced strain on the banking sector. Another downturn followed in 2011-–2012 during the Eurozone sovereign debt crisis, when spillovers from Southern Europe generated volatility despite relatively resilient Nordic fundamentals. In 2015--2016, financial markets experienced heightened volatility amid the China growth scare, the continued oil-price collapse and elevated geopolitical uncertainty in Europe associated with the refugee crisis and the Brexit referendum. In early 2020, the Covid-19 pandemic caused rapid drawdowns before an equally swift recovery supported by large-scale fiscal and monetary interventions. In 2022, the Nordic markets were hit by overlapping shocks. Firstly, russian illegal invasion of Ukraine triggered an energy and inflation surge; central banks raised rates aggressively, making equities and currencies highly volatile. Secondly, Sweden’s real estate downturn began, with housing prices and property firms under pressure, creating crisis-like conditions across the region. In early 2025, the United States implemented a new round of tariff increases affecting a broad set of trading partners, contributing to weaker global trade conditions and elevated policy uncertainty. This trade-policy shock may have weighed on the Nordic recovery by raising uncertainty for export-oriented firms and complicating investment and supply-chain decisions.

The above evidence suggests that the proposed eigenvalue-ratio indicator $\chi(t)$ provides an informative characterization of market stress episodes. Later on in this section, I apply $\chi(t)$ at the individual-market level to identify market regimes and adapt portfolio allocations accordingly.

\subsection{Portfolio performance}

\begin{figure*}[t]
	\centering
	\subcaptionbox{OMX Stockholm 30.\label{fig:plot_returns_omxs30}}[0.35\linewidth]
	{\centering
		{\includegraphics[height=3.6cm, trim={0.4cm 0cm 1.3cm 0cm}, clip=true]{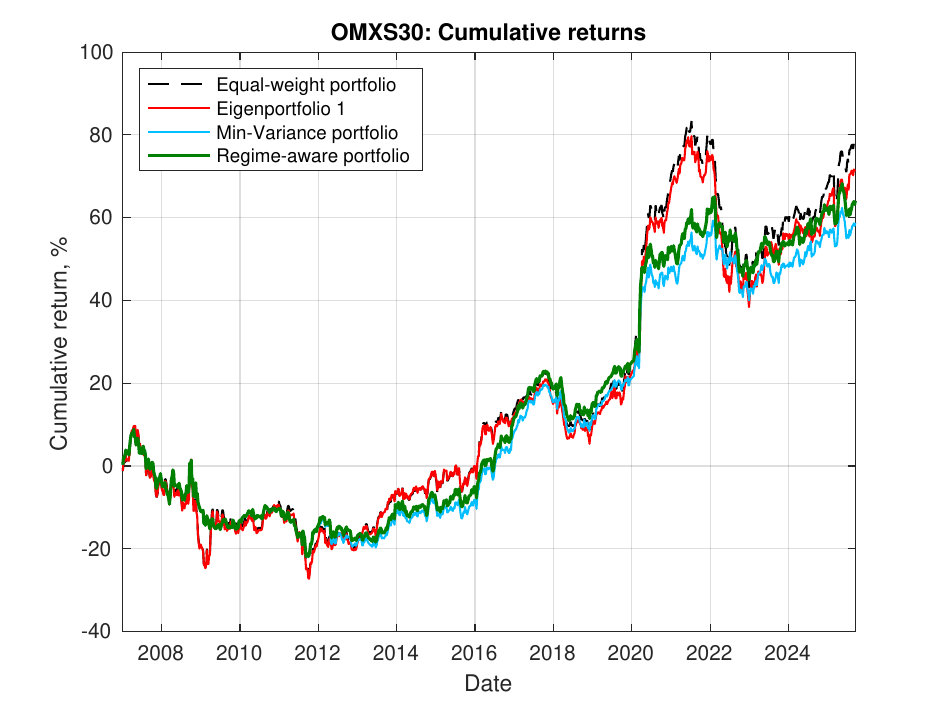}}
		}%
	\subcaptionbox{OMX Copenhagen 20.\label{fig:plot_returns_omxc20}}[0.31\linewidth]
	{\centering
		{\includegraphics[height=3.6cm, trim={1.15cm 0cm 1cm 0cm}, clip=true]{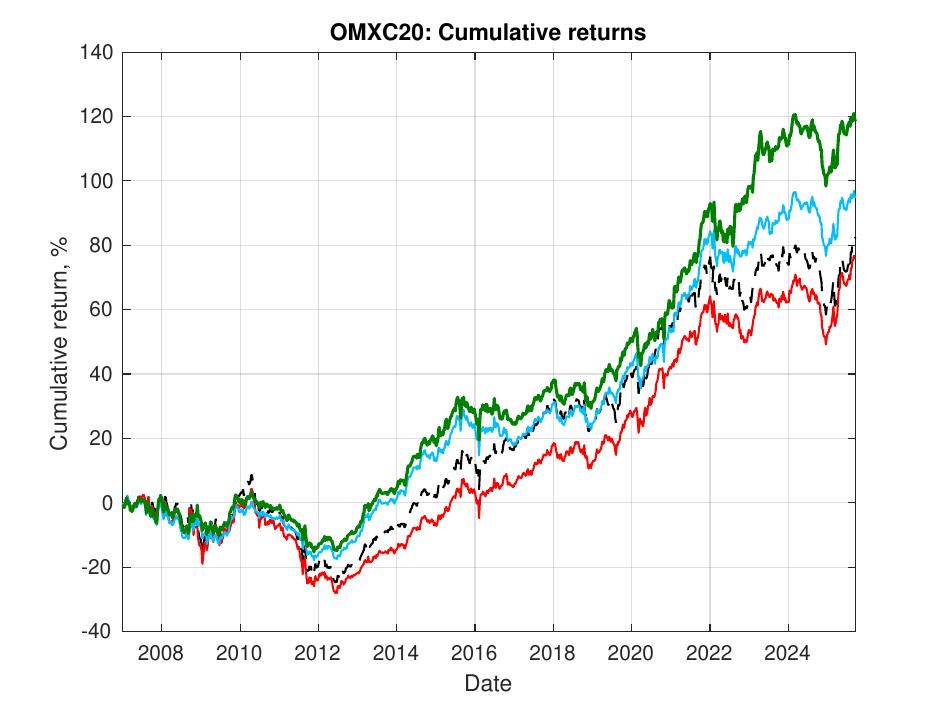}}
		}%
    \subcaptionbox{OMX Helsinki 25.\label{fig:plot_returns_omxh25}}[0.33\linewidth]
	{\centering
		{\includegraphics[height=3.6cm, trim={1.15cm 0cm 1cm 0cm}, clip=true]{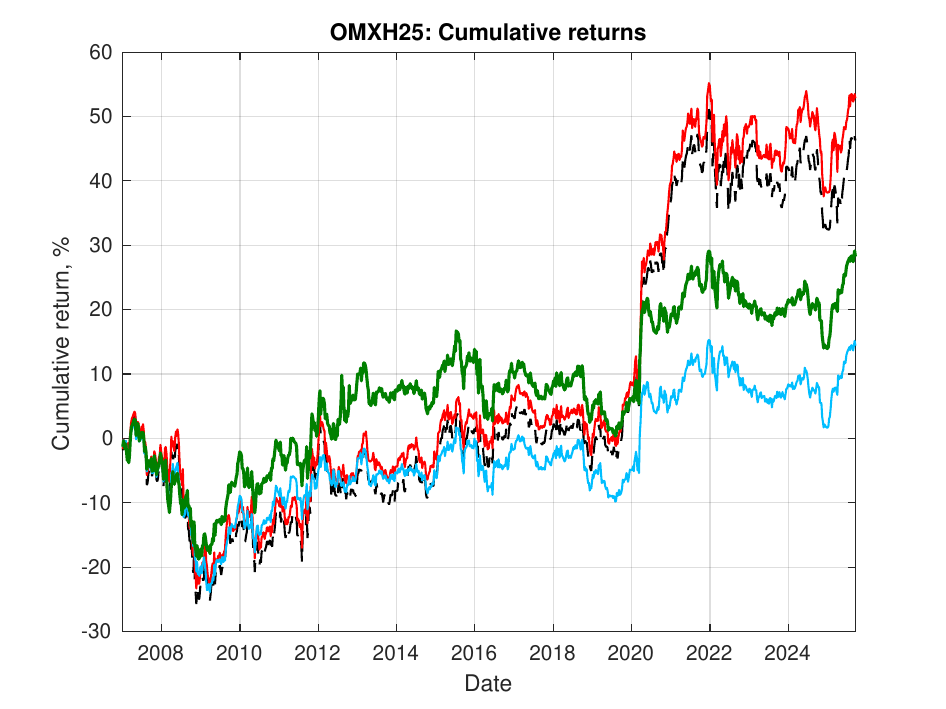}}
		}
	\caption{Cumulative-return performance of various portfolio allocations.}
	\label{fig:eigenportfolio_returns}
\end{figure*}

I next report backtesting results, showing the profit-loss (P\&L) performance of the proposed regime-aware portfolio. Figure~\ref{fig:eigenportfolio_returns} displays cumulative returns for four portfolio constructions across the three Nordic equity markets. The EW portfolio (black dashed line), computed via~\eqref{eqn:ew_portfolio}, serves as a standard baseline. As documented by~\citet{demiguel2009optimal}, the EW rule constitutes a competitive out-of-sample benchmark relative to many optimized allocations.

As a second benchmark, I include the principal eigenportfolio associated with the largest eigenvalue $\lambda_1$ of the correlation matrix (red line), constructed using~\eqref{eqn:eigenportfolio}. I also report the Markowitz MV portfolio (cyan line) obtained from~\eqref{eqn:mv_portfolio} using a denoised covariance estimate based on eigenvalue clipping~\eqref{eqn:lambdas_clipped}, further shrunk toward a constant-covariance target~\eqref{eqn:const_cov_target} with shrinkage intensity $\delta=0.1$. This MV allocation is a capable and widely used benchmark, focusing on risk minimization and not requiring expected-return forecasts.

Finally, the proposed regime-aware strategy (green line) is obtained by solving the quadratic program in~\eqref{eqn:regime_aware_problem}. I impose a ceiling of $\gamma_1=0.3$ on the portfolio exposure to the market mode and a floor of $\gamma_2=0.2$ on exposure to the second eigenmode. In stress windows identified by the eigenvalue-ratio indicator~\eqref{eqn:crisis_indicator}, the allocation is computed using the denoised and shrunk covariance estimate; in tranquil periods, I revert to the aforementioned MV rule, as described in Section~\ref{sec:portfolio_construction}.

Figure~\ref{fig:eigenportfolio_returns} reports the cumulative return trajectories of the benchmark portfolios and the proposed regime-aware strategy (green line). Across all three Nordic markets, the regime-aware strategy delivers higher cumulative returns over the sample than its uninformed MV counterpart. In two of the three markets (OMXS3 and OMXC20), the performance curves exhibit broadly similar dynamics, with the proposed adaptive strategy outperforming others in terms of risk-adjusted returns. However, as Figure~\ref{fig:plot_returns_omxh25} indicates, for the OMXH25 market the na\"{i}ve MV portfolio underperforms markedly after 2020 and does not completely recover thereafter. This pattern is consistent with elevated cross-asset correlations in that market following the Covid-19 shock, which increases the dominance of the market mode in the empirical correlation structure and can lead the MV allocation to become overly conservative. By imposing additional regime-dependent constraints during stress periods, the proposed adaptive strategy manages to improve upon the na\"{i}ve MV benchmark in OMXH25; nevertheless, it remains below the performance of the EW allocation and the principal eigenportfolio, which benefit from strong exposure to the dominant systematic component driving most of the return variation.

While cumulative P\&L curves provide an intuitive visualization of performance over time, Sharpe, Sortino and Treynor ratios offer standardized, risk-adjusted summaries that enable more rigorous and comparable evaluation of portfolios by accounting for total, downside and systematic risk, respectively. For instance, a strategy may appear attractive in cumulative terms while exhibiting significantly higher risk, or vice versa. For this reason, I provide Table~\ref{tab:summary_data} that summarizes the main performance and risk statistics of portfolios under consideration. It is noteworthy that the Treynor ratio herein is computed using the EW portfolio of the investment universe as the market proxy. Under this definition, this ratio for the EW portfolio coincides with its own arithmetic average return (i.e., $\beta$ = 1 by construction) and serves as a natural reference level for comparing alternative strategies.

The table shows that the proposed regime-aware portfolio achieves not only strong cumulative growth but also superior risk-adjusted performance relative to the na\"{i}ve MV benchmark. Consistent with Figure~\ref{fig:eigenportfolio_returns}, the adaptive strategy outperforms the state-of-the-art methods in OMXS30 and OMXC20, whereas in the Finnish OMXH25 market the EW benchmark and the principal eigenportfolio exhibit particularly strong performance. 

Overall, these quantitative results illustrate the merits of the proposed framework. They also indicate that incorporating regime information can enhance defensive positioning and stabilize allocations while preserving participation in subsequent market recoveries.

\begin{table}
    \centering
    \caption{Performance comparison between various portfolio strategies.}
    \label{tab:summary_performance}
    \begin{tabular}{lrrrrr}
        \hline
     \textbf{Portfolio}  & \textbf{Mean (ann.), \%} & \textbf{Vol (ann.), \%} &  \textbf{Sharpe} & \textbf{Sortino} & \textbf{Treynor, \%}\\
        \hline
    \textbf{OMXS30}     & &   & &     &     \\
Equal-weight           & \textbf{3.242}        & 9.416        & 0.344        & 0.491        & 3.633        \\
1st eigenportfolio     & 3.029        & 9.511        & 0.318        & 0.454        & 3.413        \\
Min-variance           & 2.590        & \textbf{7.424}        & 0.349        & 0.486        & 4.376        \\
Regime-aware           & 2.776        & 7.466        & \textbf{0.372}        & 0\textbf{.529}        & \textbf{4.671}        \\
        \hline
    \textbf{OMXC20}     & &   & &     &     \\ 
Equal-weight           & 3.394        & 8.763        & 0.387        & 0.538        & 3.723        \\
1st eigenportfolio     & 3.188        & 8.725        & 0.365        & 0.500        & 3.589        \\
Min-variance           & 3.768        & \textbf{7.153}        & 0.527        & 0.791        & 5.771        \\
Regime-aware           & \textbf{4.434}        & 7.417        & \textbf{0.598}        & \textbf{0.932}        & \textbf{6.678}        \\
        \hline
    \textbf{OMXH25}     & &   & &     &     \\ 
Equal-weight           & 2.121        & 9.538        & 0.222        & 0.328        & 2.553        \\
1st eigenportfolio     & \textbf{2.369}        & 9.353        & \textbf{0.253}        & \textbf{0.373}        & \textbf{2.845}        \\
Min-variance           & 0.745        & \textbf{7.374}        & 0.101        & 0.149        & 1.599        \\
Regime-aware           & 1.383        & 7.968        & 0.174        & 0.256        & 2.466        \\
        \hline
    \end{tabular}
    Note: Best-performing schemes per metric are highlighted in bold font.
\end{table}

\section{Conclusions}
\label{sec:conclusions}
This paper has investigated the time-varying correlation dynamics of stock returns in Nordic equity markets and assesses whether regime-dependent dependence structure can be leveraged to improve portfolio performance, particularly during episodes of market stress. It was shown that shifts in the spectral structure of the correlation matrix, most notably in the behavior of the leading and subleading eigenvalues, provide informative signals of crisis conditions. These signals can be incorporated into portfolio construction to support more defensive allocations when market-wide co-movement intensifies.

I have documented the time evolution of rolling correlation matrices and their 
eigenvalues for the OMXS30, OMXC20 and OMXH25 asset universes over a two-decade sample. Using a 
rolling-window framework, I have found pronounced regime dependence: stress episodes coincide 
with sharp increases in the leading eigenvalue and counter-cyclical dynamics in the second 
eigenvalue, consistent with the previously reported defensive role for the 
subleading mode~\citep{molero2025random}. Complementary eigenportfolio 
regressions have shown that the principal eigenportfolio exhibits a slope $\beta\approx 1$ and is 
statistically significant at the 1\% level, supporting an interpretation of the dominant eigenmode 
as a market factor in the sense of Sharpe~\cite{sharpe1964capital}.

Building on these findings, I have proposed a regime-aware portfolio strategy that conditions 
allocations on the estimated market state by integrating covariance-matrix cleaning, 
an eigenvalue-based regime indicator and a long-only minimum-variance optimization 
augmented with constraints that bound exposure to dominant eigenmodes. 
The empirical results indicate that the proposed adaptive approach 
improves downside protection and risk-adjusted performance during crisis
regimes, while remaining competitive with state-of-the-art benchmarks in tranquil
periods. 

Several directions for future research emerge from this work. First, the empirical scope 
could be broadened to include additional Nordic and European equity universes, as well as 
other asset classes, in order to assess the robustness and portability of the proposed 
regime-aware strategy. Second, from a methodological standpoint, it would be informative 
to investigate alternative regime-identification approaches, including Markov-switching models, 
dynamic conditional correlation specifications and machine-learning classifiers that combine 
spectral features with macro-financial covariates. Finally, further progress may be achieved 
by incorporating heavy-tailed and non-Gaussian extensions of the Marchenko--Pastur law, 
enabling a more realistic treatment of extreme events in eigenvalue-based filtering.

\backmatter





\bmhead{Acknowledgements}

The author gratefully acknowledges Asst. Prof. Diogo Mendes and Prof. Robert \"{O}stling for valuable guidance in finance and economics that helped shape the foundations of this study. The author also thanks Assoc. Prof. Riccardo Sabbatucci and Dr. Laura Molero Gonz\'{a}lez for helpful discussions on the application of random matrix theory in finance.


\section*{Declarations}

\noindent\textbf{Funding:} The author gratefully acknowledges financial support from the SSE Corporate Partner Scholarship 2023, which funded his participation in the education programme forming the basis of this work.

\noindent\textbf{Conflict of interest:} The author declares that they have no known competing financial interests or personal relationships that could have appeared to
influence the work reported in this article.

\noindent\textbf{Data availability:} The data associated with this publication is available with the author.

\bibliography{sn-bibliography}

\end{document}